# Implementation of Google Assistant & Amazon Alexa on Raspberry Pi

Shailesh D. Arya[1]
*Dept. of Computer Science & Engineering*
*Pandit Deendayal Petroleum University*
*Gandhinagar, Gujarat*
shailesh.ace16@sot.pdpu.ac.in

Dr. Samir Patel[2]
*Dept. of Computer Science & Engineering*
*Pandit Deendayal Petroleum University*
*Gandhinagar, Gujarat*
Samir.Patel@sot.pdpu.ac.in

*Abstract*— **This paper investigates the implementation of voice-enabled Google Assistant and Amazon Alexa on Raspberry Pi – 3. Virtual Assistants are being a new trend in how we interact or do computations with physical devices. A voice-enabled system essentially means a system that processes voice as an input, decodes, or understands the meaning of that input and generates an appropriate voice output. In this paper, we are developing a smart speaker prototype that has the functionalities of both (i.e. google Assistant and Amazon Alexa) in the same Raspberry Pi. Users can invoke a virtual assistant by saying the hot words and can leverage the best services of both eco-systems. This paper also explains the complex architecture of Google Assistant and Amazon Alexa and the working of both assistants as well. Later, this system can be used to control the smart home IoT devices (i.e. Smart Plug, Wi-Fi enabled Switch, Smart Bulb, etc.).**

## I. INTRODUCTION

In today's digital world, people are being attracted by the term 'Personal Virtual Assistant'. We have seen major shifts in our computing or interaction with the devices. Every decade or so we see a drastic change on how we interact with physical devices. We have seen Mainframe computers (1977), Desktops (1987), Internet (1997), Mobile Computing (2007). And now we shift this computing trend to Assistants (2017). With each shift in computing, we encounter lots of new devices, new ways to consume the content, interact with services through software and applications. Personal Assistants are new way of interaction with smart IoT devices in which user can ask or give command to smart device (i.e. smart speaker). Voice enabled assistants are devices that can respond to multiple voices, regardless of accent, can execute several commands or can provide an answer, thus imitating a natural conversation [1-2]. Smart device will understand the user's request with Natural Language processing, decodes the intention and generates an answer which flows back to user in Natural Language. There are some open source software packages that allow speech recognition such as Kaldi [6] or Pocket Sphinx [7]. With Assistants being new in this line-up, developers are trying to develop Personal Assistant which can act as a companion to its user, understand user's request and reply with most appropriate answer in natural language which doesn't sound robotic. Sometime, the main issue with the assistant is, user all around the globe speak different languages, with different pronunciation of words which can mislead the assistant and it can come up with irrelevant answer that can annoy the user. But with recent advancements in Machine learning and Natural Language Understanding, these Personal Assistants are becoming more and more efficient in understanding the natural language and providing relevant information. All the tech giant companies coming up with their own version of the personal assistants which provides goods and services to its users in a way that user need to do a very little work. Companies like Google, Amazon, Microsoft, Samsung etc. are constantly upgrading their version of Personal Assistants by broadening the developers and consumer space [4]. Personal Assistant are available on devices like smart phones, smart televisions, headphones, smart watches and smart speakers as well. They can also be used for Home Automation. People having disabilities can be benefited with such Personal Assistant Devices [5].

At Present time, we are surrounded with many voice-enabled Personal Assistants (i.e. Google's Assistant, Amazon's Alexa, Microsoft's Cortana, Samsung's Bixby, Apple's Siri etc.). Each of which provides similar set of services and features with extensions which can be leveraged by users to control their way of life.

Our goal for this paper is to develop a Voice enabled Smart Speaker Prototype with Raspberry Pi providing services offered by Google Assistant and Amazon Alexa in same piece of hardware [8]. It will also work for Home Automation System for controlling Smart Devices with Assistants. User can use both services alternately by invoking them with hot words (i.e. Ok google, Alexa). This model works on primary input of user's voice. Any person ranging from child to old-age person can use this Smart Speaker using Raspberry Pi to get answers for usual queries, enjoy entertainment, schedule day, manage tasks, reminders, control smart home etc. The gadget is fit for voice conversations, music playback, making plan for the day, setting alarms, podcast streaming, playing audiobooks, and giving climate reports, traffic and other information.

The following is the set of equipment used for this project.

Raspberry Pi

Micro SD Card

Power Supply

USB Microphone

Speaker

Ethernet Network Connection or Wi-Fi

For this project, the model used for RPi is Raspberry Pi 3 Model B+ with Raspbian OS installed. RPi is connected with local Wi-Fi network. USB Microphone is connected to RPi using USB socket and speaker is connected to RPi using 3.5 mm AUX cable. The system requires a constant power source to run the Raspberry Pi and an Internet connection either through Ethernet cable or wi-fi is also required to process the use's command on cloud.

## II. PROPOSED WORK

The system comprises the equipment list mentioned above. Initially, when the user stars the Raspberry PI, both the program for Google Assistant SDK and Amazon Alexa Service will run automatically waiting for the hot keyword used to invoke them.

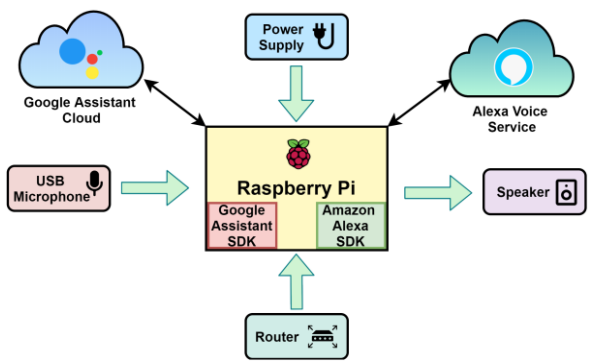

Fig.1. Overall System Architecture

User uses microphone attached to Raspberry Pi to initiate the conversation. Raspberry Pi has limited memory, storage and resources, hence running a full machine learning model directly on Raspberry Pi is not feasible. Hence, User's query will be recorded and sent to Google's Cloud service where the Machine Learning Model for Natural language Understanding has been implemented. It extracts the intent of the query by performing all NLP tasks (i.e. Tokenization, parsing, Semantic analysis, information extraction, intent etc.). After the intent is identified, Cloud service will generate the results for the matched intent which will go through the dialogue generation phase and sent to Raspberry Pi. Finally, User can hear the voice result in natural Language from the speaker attached to the Raspberry Pi.

Similarly, Amazon's Alexa services also works on same foot print. It detects the hot word (i.e. 'Alexa') for the microphone attached to RPi and sends the user query to its cloud function where Machine learning model is implemented. Tasks like Query parsing, intent extraction, fetching the related result for the matched intent and returning the results are done on cloud and at the end user can get the results on RPi in Natural Language.

In this system, Raspberry Pi requires 5v/2amp constant power supply. Also, it'll require a healthy internet connection either wired or wireless (i.e. Wi-Fi). Router is used for forwarding the data packets or information between the network s or within a network or to another network. In this system, some other devices are also used such as Wi-Fi enabled Smart socket and Smart Switch for Home Automation. Smart Socket is used to connect small electronic devices (i.e. Charger, Smart bulbs, Fairy lights, night lamps etc.) Wi-Fi enabled Smart Switch is used to control heavy electronic devices like Fans, Tube lights, Ac etc.

### A. Working of Google Assistant

Key Terms: Action, Intent and Fulfilment

**Action:** Action is an entry point to start conversation with Google Assistant.

To start conversation with Google Assistant, User can say the hot word, along with the action name (i.e. "Hey Google! Talk to BMI Estimator")

**Intent:** The task that user wants to perform with the Assistant.

Task can be as simple as searching word on web, to complex tasks like controlling smart devices with Assistant.

**Fulfilment:** A service, App or logic that handles an intent and carries out corresponding action.

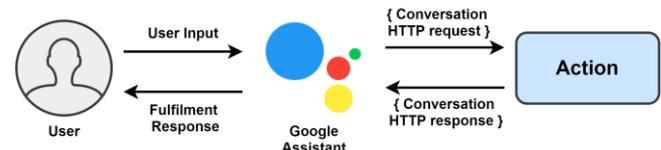

Fig.2. Working of Google Assistant

From this point onwards, user's conversation with Assistant will be bi-directional until user's intent is fulfilled.

Behind the scenes during conversation:

- Your action run entirely on cloud, even when invoked with different devices.
- Every Action has particular intents and corresponding fulfilment for that Intent.
- The user's device sends the user's utterance to the Google Assistant, which routes it to your fulfilment service via HTTP POST requests.
- Your fulfilment figures out a relevant response and sends that back to the Assistant, which ultimately returns it to the user.

### B. Working of Amazon Alexa

The following figure-3 the workflow of how Alexa works.

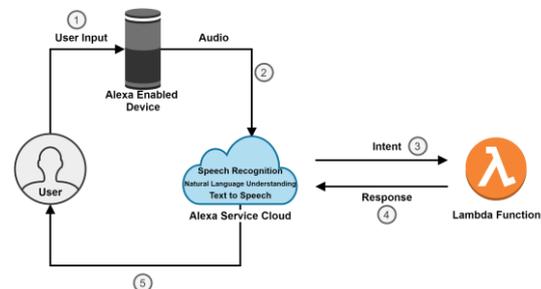

Fig.3. Working of Amazon Alexa

Amazon Alexa works on similar footprint of Google Assistant.

- To start conversation with Alexa enabled device, user says the hot key word ("Alex – with some query").
- The Alexa-enabled device sends the utterance to the Alexa service in cloud. There, the utterance is processed via automatic speech recognition, for conversion to text, and natural language understanding to recognize the intent of the text.

- Alexa service sends a JavaScript Object Notation (JSON) request to an AWS Lambda Function in the cloud to deal with the intent. The Lambda function works as the backend and executes code to deal with the intent and returns the outcomes back to the Alexa enabled device.

### III. IMPLEMENTATION

This section will guide you on how to install Google Assistant SDK and Amazon Alexa Service on Raspberry Pi. For this project, I have used Raspberry Pi 3 Model B with Raspbian OS. You can download and Install the OS from their Official Site. It is better to use a 16 GB micro SD card for the installation purpose.

A. Installing Google Assistant SDK on RPi

The following are the Steps for integrating Google Assistant SDK in Raspberry Pi

1. Set Up Hardware and Network Access
2. Configure Speaker and Microphone
3. Create Project on actions.Google Console & Register the Device Model
4. Install the SDK and Sample Code
5. Run the Sample Code

1. Setup Hardware:
   Component list:
   1. Raspberry Pi 3 model B and Power Supply

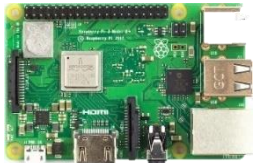

   2. USB Microphone

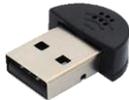

   3. Speaker (with 3.5 mm headphone jack)

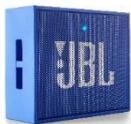

   4. SD card with Raspbian OS installed

2. Configure Speaker and Microphone:
   Make sure the USB Microphone is connected to RPI through USB Socket and Speaker is connected through AUX cable to Raspberry Pi. On RPi, by default audio output is through HDMI so we need to change it to 3.5 mm jack. Find your recording and playback device connected to RPI y typing the following command in terminal:

- $ arecord -l (locate USB Microphone in the list)

- $ aplay -l (Locate you speaker in the list. Note that 3.5 mm-jack is typically labelled as analog or bcm2835 ALSA)

Note down the card number and device number. (i.e. for USB Mic: Card 1, Device 0). Next, create a new file named as **.asoundrc** in /home/pi directory and put the following content in it. Also replace card number and device number as well.

```
pcm.!default {
  type asym
  capture.pcm "mic"
  playback.pcm "speaker"
}
pcm.mic {
  type plug
  slave {
    pcm "hw:<card number>,<device number>"
  }
}
pcm.speaker {
  type plug
  slave {
    pcm "hw:<card number>,<device number>"
  }
}
```

You can verify your installation by typing following commands in terminal to check speaker and microphone works correctly.

- $ speaker-test -t wav (Plays sound from speaker)

- arecord --format=S16_LE --duration=5 --rate=16000 --file-type=raw out.raw (Record your audio and playback)

3. Create a Project on actions.google console:
   3.1. Go to actions on google site and sign in with your google account.
   3.2. Click on add/import project.
   3.3. Give your project a Name (i.e. gBot), select language and country.
   3.4. Go to Device registration on the bottom of the page.
   3.5. Register your model and download the OAuth credentials. Keep the credentials in the home directory and do not rename it.
   3.6 Enable Google Assistant Api for the registered device from this link: https://bit.ly/2TXa1zV

4. Install the Google Assistant SDK on RPi:
Run the following set of command one by one in terminal.

- sudo apt-get update
- sudo apt-get upgrade
- sudo apt-get install python3-dev python3-venv
- python3 -m venv env
- env/bin/python -m pip install --upgrade pip setuptools wheel
- source env/bin/activate
- sudo apt-get install portaudio19-dev libffi-dev libssl-dev libmpg123-dev
- python -m pip install --upgrade google-assistant-library
- python -m pip install --upgrade google-assistant-sdk[samples]
- python -m pip install --upgrade google-auth-oauthlib[tool]

This will take some time so be patient. After the installation is complete, we need to authorize the Raspberry Pi Device inorer to use Google Assistant services. Make sure OAuth Configuration file (JSON file) is in /home/pi directory. Run the following command with changed client_secret_XXXX in the end.

- google-oauthlib-tool --scope https://www.googleapis.com/auth/assistant-sdk-prototype \

  --scope https://www.googleapis.com/auth/gcm \

  --save --headless --client-secrets client_secret_XXXXXX.json

This will prompt a URL in terminal, open that URL and authenticate the device by signing in with registered email. Finally, you need to paste the code in terminal for authorization.

5. Run Google Assistant on RPi:

First activate the python3 virtual environment
- source /home/pi/env/bin/activate

Run the following command to start interaction with google assistant using hot word (i.e. ok Google!)
- googlesamples-assistant-hotword  --device-model-id <model id>

B. Installing Amazon Alexa Service on RPi

Note: Step 1 and 2 are identical with the steps in installing Google Assistant on RPi mentioned earlier.

3. Register yourself on Amazon Developer.

4. Device Registration:

Go to Alexa Voice Service and then to products. Fill the required details (i.e. Product name, ID, Product category etc.). Select Device with Alexa Build in. Next, set up a new Security Profile and download the JSON file from other devices and platform tab. This file will be used to authenticate use while installing Alexa services.

5. Installing Alexa services on RPi

Create a new folder named as Alexa in /home/pi directory and paste the JSON file downloaded earlier in this folder. Now, open Terminal in the same directory and run the following commands one by one.

- sudo apt-get update
- sudo apt-get upgrade
- wget https://raw.githubusercontent.com/alexa/avs-device-sdk/master/tools/Install/setup.sh \
- wget https://raw.githubusercontent.com/alexa/avs-device-sdk/master/tools/Install/genConfig.sh \
- wget https://raw.githubusercontent.com/alexa/avs-device-sdk/master/tools/Install/pi.sh
- sudo bash setup.sh config.json [-s 1234]\

(Note: When Asked to agree the terms and conditions in terminal, just type 'AGREE')

Once you see the text 'Completed Configuration/Build' on terminal you can go to next step.

6. Get refresh Token:

To grant Alexa to control device we required to opt for a token. To do the same type the following command in the terminal.

- sudo bash startsample.sh

Copy the code to the link: amazon.com/us/code mentioned in terminal. Allow terms and conditions and lastly a screen will pop up wih registration successful message.

Fig.4. Alexa Device Registration

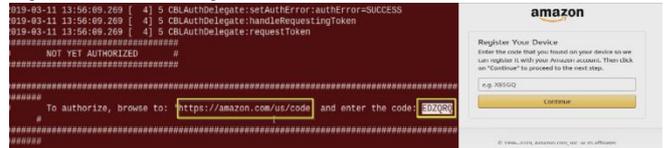

7. Run Alexa:

Go to /home /pi/alexa directory and type the following command in terminal:

- $ sudo bash startsample.sh

Alexa services will start and you can start conversation by saying the hot word (i.e. 'Alexa').

The final Hardware aesthetic for this project looks as follow:

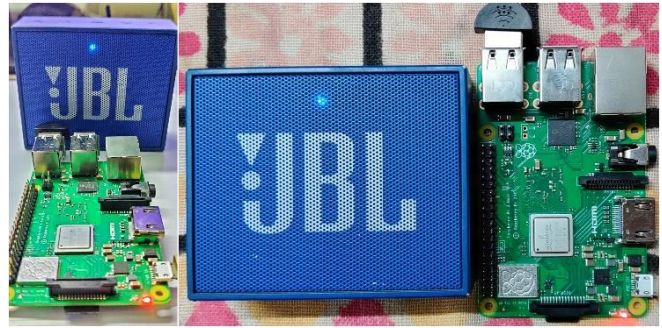

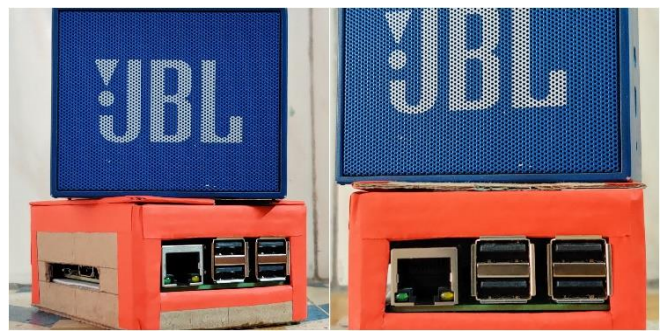

Fig.5. Smart Speaker Prototype

## IV. Conclusion

This paper introduced the possibility of using Google Assistant and Amazon Alexa on same Raspberry Pi microcontroller. Starting from scratch, one can come up with a pretty good Smart Speaker prototype which has both assistants built in and can be used in daily basis. Smart speakers like Google home, Amazon echo and many other 3[rd] party smart speakers are available in market but with functionality limited to any one of the assistant. Maybe the major reason that vendors do not include both assistant in the same device is to not create confusion for the users. The work shown in this paper is a good DIY project for any tech enthusiast and one can get a smart speaker for the money less than any consumer product available. This speaker can be used with any IoT appliances (i.e. smart switch, smart bulb, smart socket) that supports Google Assistant or Amazon Alexa integration to control devices at home.

## REFERENCES

[1] P. Milhorat, S. Schogl, G. Chollet et all "Building the Next Generation of Personal Digital Assistants" 1st International Conference on Advanced Technologies for Signal and Image Processing – ATSIP'2014, March 17-19, 2014, Sousse, Tunisia, pp.458–463.

[2] V. Kepuska and G. Bohouta, "Next Generation of Virtulal Personal Assistants (Microsoft Cortana, Apple Siri, Amazon Alexa and Google Home)", 2018 IEEE 8th Annual Computing and Communication Workshop and Conference 8-10 Jan. 2018 Las Vegas, USA, pp.99–103.

[3] Achal S Kaundinya, Nikhil S P Atreyas, Smrithi Srinivas, Vidya Kehav , Naveen Kumar M R. "Voice Enabled Home Automation Using Amazon Echo", International Research Journal of Engineering and Technology (IRJET), Aug-2017.

[4] Dhiraj S. Kalyankar, Prof. Dr. P. L. Ramteke, "Personal Google API Assistant System using Raspberry Pi", International Research Journal of Engineering and Technology (IRJET), Feb 2019.

[5] Komal P. Nikure , Prof.Ashish Manusmare, Review of Dynamic Digital Assistant Using Raspberry Pi, IOSR Journal of Electronics and Communication Engineering (IOSR-JECE), Vol. 14, Issue 3, pp 36-39.

[6] Kaldi Toolkit for Speech Recognition, http://kaldi-asr.org/index.html, accesssed July 7, 2018.

[7] Open Source Speech Recognition Toolkit, https://cmusphinx.github.io/, accessed July 7, 2018.

[8] Septimiu Mischie, Liliana Matiu-Iovan, Gabriel Gasparesc "Implementation of Google Assistant on Rasberry Pi" International Symposium on Electronics and Telecommunications (ISETC), November 2018.